\documentclass[aps,prb,twocolumn,showpacs]{revtex4}

\usepackage{graphicx}
\DeclareGraphicsExtensions{.png,.jpg,.eps}
\usepackage{braket}
\usepackage{xcolor}
\usepackage{amsmath}

\begin{document}

\title{Controlled quantum state transfer in $XX$  spin chains at  the 
Quantum Speed Limit}

\author{ D. S. Acosta Coden$^{1}$, S. S.  G\'omez$^{1}$, 
A. Ferr\'on$^{1}$,   O. Osenda$^{2}$
}
\affiliation{ 
(1) Instituto de Modelado e Innovaci\'on Tecnol\'ogica (CONICET-UNNE) and 
Facultad de Ciencias Exactas, Naturales y Agrimensura, Universidad Nacional 
del Nordeste, Avenida Libertad 5400, W3404AAS Corrientes, Argentina.
\\(2) Instituto de F\'sica Enrique Gaviola, Universidad Nacional de C\'ordoba
, CONICET, Facultad de Matem\'atica, Astronom\'ia, F\'isica y Computaci\'on,
Av. Medina Allende s/n, Ciudad Universitaria, CP:X5000HUA C\'ordoba, 
Argentina 
}

\date{\today}

\begin{abstract}

The Quantum Speed Limit can be found in many different situations, in 
particular in the propagation of information through quantum spin chains. In 
homogeneous chains it implies that taking information from one extreme of the 
chain to the other  will take a time  $O(N/2)$, where $N$ is the chain 
length. Using Optimal Control Theory we design control pulses that achieve near 
perfect population transfer between the extremes of the chain at times on the 
order of $N/2$, or larger, depending on  which features of the transfer 
process are to be studied. Our results show that the control pulses that 
govern the dynamical behaviour of chains with different lengths are closely 
related, that larger control times imply more complicated control pulses than 
those found at times on the order of $N/2$ and also larger driving energies. 
The 
pulses were constructed for control schemes involving one or two actuators in 
chains with exchange couplings without static disorder. Our results also show 
that the two actuator scheme is considerably more robust against the presence 
of static disorder than the scheme that uses just a single one.   

\end{abstract}
%\pacs{73.22.Pr, 73.43.Cd}

\maketitle

\section{Introduction}\label{sec:introduction}

The transfer, or transmission, of quantum states in different communication 
channels has been touted as a task of paramount importance to routinely
implement  Quantum Information Processing (QIP) and Technologies. Since the 
paper of Sougato Bose in 2003 \cite{Bose2003} the area has grow enormously and 
the understanding of the field has been collected in books 
\cite{Nikolopoulos2015} and other excellent reviews \cite{Bose-review}. Also 
from the beginning, it was understood that the severe restrictions imposed by 
the fragility of the quantum states to be transferred required {\em perfect} or 
near perfect quantum transmission so the errors  do not spoil the state for 
further use after it has been transmitted.

The transmission channel is constructed from copies of a given physical system. 
The copies are spatially distributed along a line or in other spatial 
distributions \cite{Kostak2007} and might or might not interact with a 
surrounding environment. 
Many different physical systems have been considered as the basic unit of the 
communication channel, quantum dots \cite{quantum-dot-chain}, superconductor 
qubits \cite{Li2018}, nuclear spins 
\cite{nuclear-spin-chain}, cold atoms or ions trapped in  arrays 
\cite{Loft2011,Banchi2011prl} or coupled waveguides \cite{Chapman2016}. Anyway, 
most 
theoretical studies focus on Heisenberg or 
$XY$ spin chain models since these models captures the basic physical traits 
that are of interest. In the case of quantum dots chains most theoretical 
studies deal with the Hubbard model \cite{Hawrylak} although 
other approaches are feasible \cite{Coden2019}.   

In models without environment, {\em i.e.} with unitary dynamics, there are two 
main approaches to achieve perfect or near perfect sate transfer. In the 
simpler approach the Hamiltonian unitary dynamics takes the state to be 
transferred from one extreme of the spin chain to the other and the 
spin chain Hamiltonian is time-independent. Several possibilities have been 
studied, for example, chains that achieve perfect transmission by engineering 
all the spin couplings to special values 
\cite{Christandl2004,Christandl2005,Yung2006}, conclusive and 
perfect transmission using to parallel spin chains 
\cite{Burgarth2005,Burgarth2005b}, boundary 
controlled spin chains \cite{Zwick1,Zwick2,Zwick3}, optimal dynamics 
\cite{Banchi2010,Banchi2011}, and so on. The second approach 
assumes that the system will be driven by {\em control} fields, so the 
dynamical behavior is governed by the free Hamiltonian and a number of external 
fields. Both approaches share some features, usually the dynamics is confined 
to some sub-space of the whole Hilbert space, most commonly a sub-space with 
fixed number of excitations or magnetization. This assumption reduces 
drastically the size of the computational problem and makes feasible to study 
systems with static disorder.

Quantum spin chains of different kinds are {\em controllable} 
 in the whole Hilbert space  \cite{Jurdjevic1972,Ramakrishna1995,Burgarth2009} 
and in sub-spaces of fixed 
magnetization \cite{Wang2016} even with very few actuators, moreover 
Heisenberg spin chains are controllable with just a single actuator. There 
are numerous papers dealing with the controllability of these models using 
an external magnetic field applied to the first spin, {\em i.e.} the spin where 
the initial state is prepared to be transmitted. Actually, besides the 
transmission of quantum states there are other quantum entangling gates that 
can be controlled along a spin chain \cite{Burgarth2010}. The preparation of 
particular target states has also been considered 
\cite{Stefanatos2019,Watanabe2010}, as well as the possibility of controlling 
the state transfer applying fields to all the sites \cite{Gong2007,Murphy2010}.
Finally, it is important to mention some advances in the last five years in 
the field of Scanning Tunnel microscopy (STM). In these years we could see 
that Paramagnetic Electron Resonance (EPR) implemented using STM   
\cite{bauman2015} permits one to drive the electronic and nuclear spins of 
individual atoms on surfaces, as well as artificially created structures, such 
as dimers \cite{prr6,prr9}. In addition, STM technique allows to build longer 
chains\cite{bryant2015,spinelli2014} that can be used for control tasks by 
implementing protocols through STM-EPR.

The logic behind a reduced number of control fields, and that they are applied 
to the extremes of the chain, is based on what can be feasible on actual 
implementations, where each control field could add new sources of errors and 
the 
application  of an external field to a single element of a  truly nano scale 
device is troublesome at least, in particular it is located in the middle of 
other 
elements alike. 

Inspired by control techniques typical of Nuclear 
Magnetic Resonance, the design of control pulses based on trains of square 
pulses of varying length in time has been intensely studied in Heisenberg 
spin chains \cite{Wang2010,Heule2010}. In these studies the strength 
of the magnetic field applied to the first site of 
the chain is constant during an interval and different optimization procedures 
are  used to determine the length of the interval and the strength of the 
field. As is common in studies involving Optimal Control Theory the control 
time is fixed {\em a priori}, in the case of a single control field this time 
has been chosen on the order of $N^2$  \cite{Wang2016,Burgarth2010,Poggi2016}, 
where $N$ is the number of spins in the chain, but for more involved control 
algorithms shorter times, compatible with Quantum Speed Limit can be achieved 
\cite{Murphy2010}. 

In this paper we present strategies to control state transfer in $XX$ spin 
chains that achieve fast transfer times, $O(N)$, with very high fidelities and 
with smooth pulses that control the coupling between the first pair, or the 
first and last, pair of spins. Our results show that the shape of the pulses 
for different chain lengths are closely related, and that the external 
forcing also shows the effect of the Quantum Speed Limit, restricting the 
time interval in which it is necessary to achieve the transfer.

The paper is organized as follows. In sec. II, we present the model Hamiltonian
for the spin chain with and without static disorder and we analize how 
to design pulses in order to transfer excitations from the first site
to the last one.  In sec. III we address the possibility of controlling 
excitations using the optimal control theory in order to design a one 
actuator control strategy. Sec IV deals with the possibility of 
performing control tasks with a slightly different strategy, in this case 
we use pulses designed for two actuators. In both sections we analize the 
inclusion of static disorder and the efects in the control strategies. Finally,
in Sec V we summarize the conclusions of our work.

\section{Model Hamiltonian and Optimal Control Theory}
\label{sec:model-oct}

\subsection{Model Hamiltonian, with and without disorder and dynamics}
We consider a spin-$1/2$ chain with interactions between
nearest neighbors, described by the Hamiltonian
\begin{equation}\label{eq:xx-Hamiltonian}
	H=-\frac{\hbar}{2}\sum_{i=1}^{N-1}J_i
\left(\hat{\sigma}_i^{x}\hat{\sigma}_{i+1}^x+
	\hat{\sigma}_i^{y}\hat{\sigma}_{i+1}^y\right)
\end{equation}

\noindent where $\sigma_i$ are the Pauli matrices, $N$ is the chain length and
$J_i\geq 0$ are the exchange interaction couplings. In this work we set
$J_i=1$ for $i=2,3,...,N-1$ and $J_1=J_N=\alpha$.

We have introduced the Hamiltonian  of
a  spin chain without any external perturbation. However,
since the perfect engineering of all spin couplings is highly
improbable, it results interesting to analyze the performance of different
spin-coupling distributions related to imperfect construction of such
chains. To study the robustness of the spin chains against perturbations we
introduce static random spin-coupling imperfections quantified by $\delta_i$

\begin{equation}\label{eq:static-disorder}
J_i \rightarrow J_i (1 + \delta_i),
\end{equation}
where each $\delta_i$ is an independent uniformly distributed random
variable in the interval $[-A,A]$, this way of introducing static disorder in 
quantum spin chains has been analyzed extensively, see references 
[\onlinecite{Zwick2,Zwick3,Petrosyan2010}] and references therein. $A$ is a 
positive real number that characterizes the maximum perturbation strength
relative to $J_i$. The kind of disorder depends on the particular
experimental method used to engineer the spin chains.

For ferromagnetic couplings $J_i\geq 0$ the fundamental state is the state 
without excitations (all the spins down). The Hamiltonian in 
Eq. (\ref{eq:xx-Hamiltonian}) commutes with the total magnetization in the 
$z$ direction, so its is possible to analyze problems restricted to sub-spaces 
with fixed number of excitations, {\em i.e.} fixed magnetization. Following the 
protocol for quantum state transfer introduced by Bose \cite{Bose2003} our 
analysis will be restricted to the sub-space of one excitation. Moreover, since 
it has been shown that in this subspace the transfer fidelity averaged over 
realizations of the disorder, Eq. (\ref{eq:static-disorder}), is a function 
of the probability amplitude that an excitation in the first site of the chain 
be transferred to the last site in the next Sections we will focus in the 
``population'' of the last site, as a function of time.

Before proceeding to study particular systems it is worth to consider, once 
more, which kind of couplings are best suited to be implemented experimentally 
in order to obtain a chain easily controllable, but without too much coupling 
engineering. As has been above the engineering of all couplings to the precise 
values required to obtain perfect transmission as in References 
\cite{Christandl2004} and \cite{Christandl2005} is beyond what is feasible. 
Spin chains with only two modified couplings, with respect to a otherwise 
homogeneous spin chain, can be tuned to transfer states with high fidelity 
\cite{Banchi2010,Zwick1} and are surprisingly robust to static disorder 
\cite{Zwick2,Zwick3}, so they seem as a good option as the coupling 
distribution of choice.

\subsection{Optimal control of the state dynamics}

For controlling excitations in spin chains, we propose to design a 
time-dependent pulses able to drive an excitation in the first site 
of the chain $\Psi(0)=\Psi_1$ to an excitation in the last site of the chain 
$\Psi(T)=\Psi_N$, in a given prescribed time $T$ as short as possible to avoid 
unwanted effects of the environment over the coherence of the system.

We assume that the exchange couplings of the ends of the chain are controled 
by electric pulses and have a time-dependent shape given by the functions 
$F(t)$ and $G(t)$ for the first and last coupling respectively. Then, the 
state $\Psi(t)$ must satisfy the time-dependent Schr\"odinger equation

\begin{equation}
i\frac{\partial \Psi(t)}{\partial t}=H \Psi(t) = [H_0-\hat{h}_LF(t)-\hat{h}_RG(t)]\Psi(t).
\label{TDSE}
\end{equation}

\noindent where $H_0$ is defined in Eq. \ref{eq:xx-Hamiltonian} and

\begin{equation}
	\hat{h}_L=\frac{\hbar}{2}\left(\hat{\sigma}_1^{x}\hat{\sigma}_2^{x}+
	\hat{\sigma}_1^{y}\hat{\sigma}_2^{y}\right),	\qquad \hat{h}_R=\frac{\hbar}{2}\left(\hat{\sigma}_{N-1}^{x}\hat{\sigma}_N^{x}+
	\hat{\sigma}_{N-1}^{y}\hat{\sigma}_N^{y}\right).
\label{pert}
\end{equation}

Optimal control theory (OCT) provides a systematic method to calculate 
variationaly the optimal pulses, $F_{\rm OCT}(t)$ and $G_{\rm OCT}(t)$, that 
drives the state of the system from an initial state to an state that maximize 
the overlap with a prescribed target state $\phi_T$. We briefly outline here 
the theory and some technical details of the method as applied in this work 
\cite{werschnik2007,putaja2010}.

The total functional to be extremized is the sum of three terms, 
$J=J_1+J_2+J_3$. The first functional measures the yield of the control 
process and is defined by
%------------------------
\begin{equation}
J_1[\Psi]=|\langle\Psi(T)|\phi_F\rangle|^2,
\end{equation}
%------------------------
where $|\phi_F\rangle$ is the ``target state''.

The second functional is introduced in order to avoid high energy fields and 
is defined in term of the fluence (the time-integrated intensity of the field),
%------------------------
\begin{equation}
J_2[{\bf\varepsilon}]=-\alpha_L\left[\int_0^T F^2(t)dt\right]-\alpha_R\left[\int_0^T G^2(t)dt\right],
\end{equation}
%------------------------
where $\alpha_L$ and $\alpha_R$ are time-independent Lagrange multipliers. The 
last functional $J_3$ ensures that the electronic wavefunction evolves 
according to the time-dependent Schr\"odinger,
%--------------------
\begin{equation}
J_3[{\bf\varepsilon},\Psi,\chi]=-2Im \int_0^T \left\langle\chi(t)\left|\frac{\partial}{\partial t}-H(t)\right
|\Psi(t)\right\rangle
\end{equation}
%--------------------
where $\chi(t)$ is a time-dependent Lagrange multiplier. The variation of 
the total functional with respect to
$\Psi(t)$, $F(t)$ and $\chi(t)$ allows us to obtain the desired control 
equations \cite{werschnik2007}
%--------------------
\begin{eqnarray}\label{eq:control1}
i\frac{\partial \Psi(t)}{\partial t}&=&H
\Psi(t)\;,\; \Psi(0)=\phi,\\
\label{eq:control2}
i\frac{\partial \chi(t)}{\partial t}&=&H(t)\chi(t),\\
\label{eq:control3}
\chi(T)&=&|\phi_F\rangle\langle\phi_F|\Psi(T)\rangle,\\
\label{eq:control4}
F(t)&=&-\frac{1}{\alpha_L}Im\langle\chi(t)|\hat{h}_L|\Psi(t)\rangle,\\
G(t)&=&-\frac{1}{\alpha_R}Im\langle\chi(t)|\hat{h}_R|\Psi(t)\rangle.
\end{eqnarray}
%--------------------
This set of coupled equations can be solved iteratively. 

The solution of the system of coupled control equations can be 
obtained using several approaches, for example, using the efficient 
forward-backward propagation scheme developed in Ref. \cite{kosloff1989}. The 
algorithm starts with propagating the state of the system from $| \phi \rangle$
forward in time, using in this ``zero'' iteration a guess of the pulse fields 
$F^{(0)}(t)$ and $G^{(0)}(t)$. At the end of this step we obtain the 
wavefunction $| \Psi^{(0)}(T) \rangle$, which is used to evaluate 
$| \chi^{(0)}(T) \rangle  =|\phi_F\rangle\langle\phi_F|\Psi^{(0)}(T )\rangle$. 
The algorithm  continues with propagating $|\chi^{(0)}(t) \rangle$ backwards 
in time. At the end of this backward propagation we know both $\Psi^{(0)}$ 
and $\chi^{(0)}$ at the same time and we can obtain a first version of the 
optimized pulses $F^{(1)}(t)$ and $G^{(1)}(t)$. We repeat this operation until 
the convergence of $J$ is achieved. The values of the  Lagrange multipliers 
$\alpha_{R,L}$ are chosen in order to achieve the desired fluences. 

The 
numerical integration of the forward and backward time evolution was performed 
using high precision Runge-Kutta algorithms under various initial guesses 
(e.g., constant, random, monochromatic and a superposition of two harmonic 
pulses). The resulting optimal pulse and the time evolution under its driving 
were found to be robust for all these various starting conditions.

Note that if the target state $|\phi_F\rangle = \Psi_N$, the target probability 
is also the  ``transferred population'' after the controlled transfer protocol 
has taken place. From now on we use indistinctly both terms.

%===========================================

\begin{figure}[hbt]
\includegraphics[width=1.0\linewidth]{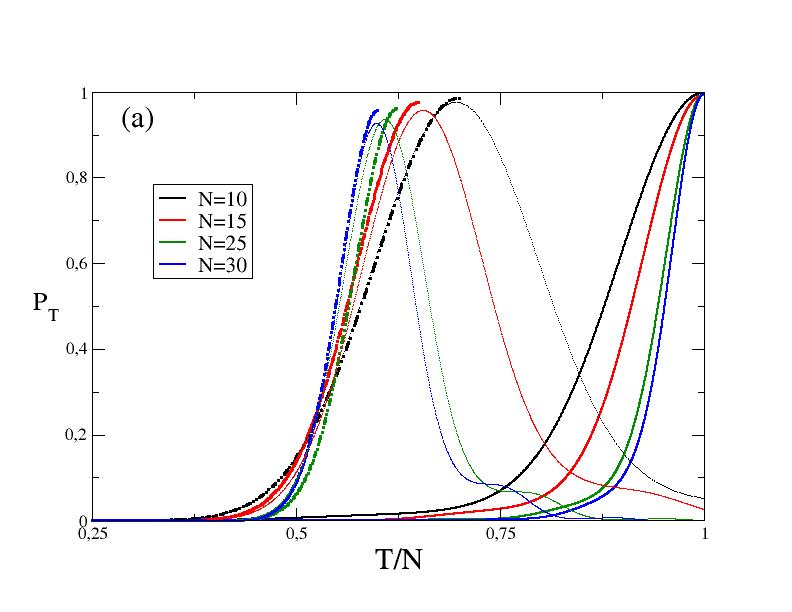}
\includegraphics[width=1.0\linewidth]{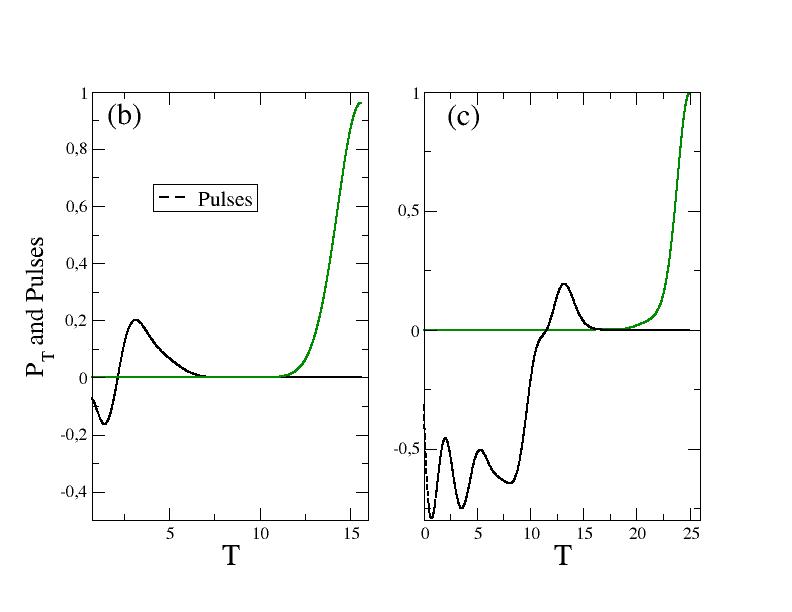}
\caption{(a) Time evolution of the target probability as a function
of the normalized time. Continuous lines show the evolotion for long
OCT pulses with $T_{OCT}=N$, dashed lines shows the evoltion for
short OCT pulses with $T_{OCT}\simeq T_{Peak}$ and dotted lines show
the free evolution. (b) Optimized pulse for short time operation 
and the target population obtained using this pulse for $N=25$. (c) Optimized 
pulse for long time operation and the target population obtained using this 
pulse for $N=25$.
}\label{fig1}
\end{figure}

\section{Control of spin excitation in a $XX$ chain:one actuator}
\label{sec:oct-one}

In this section we present results corresponding to the one actuator OCT 
protocol, as defined in Eq. (\ref{TDSE}) using $h_R=0$. The main objective of 
the present protocol is to drive an excitation from the first site to the last 
one through the manipulation of the coupling between the first and second site.
As it is known, the free evolution of the excitation, i.e. the evolution 
without external control, presents a maximum at a certain 
time which depends upon the size of the chain. In Fig. \ref{fig1} we 
compare our OCT results for two different operation times with the 
corresponding free evolution for various chain sizes. 
For most of our calculations, as in the calculations performed for 
Fig. \ref{fig1}, we will consider  boundary-controlled XX chains, and we will 
choose $\alpha$ given by the optimal value obtained for the free evolution 
\cite{Zwick1}.

In Fig. \ref{fig1} (a)  we plot the target probability, {\em i.e.} the 
population at the last site of the chain,  for the free
evolution in chains with $N=10$, $15$, $25$ and $30$ sites (dotted lines) as a 
function of a normalized time in terms of the chain size. For instance, the 
maximum probability for $N=10$ is $P_{Free}(T_{peak})=0.976$ at $T_{peak}\simeq
7 a.u.$ while for $N=30$ we have $P_{Free}(T_{peak})=0.928$ 
at $T_{peak}\simeq 18 a.u.$. In this figure we also plot our OCT results
for the same chain sizes and for two different operation times. Continuous
lines show the results for long pulses ($T_{OCT}=N$) and dashed lines
show the results using shorter pulses obtained for times where the free 
evolution has its maximum. In Fig. \ref{fig1} (b) and (c) we observe the time
evolution of the target probability for $N=25$ for both operation times. Shorter
operation times are shown in Fig. \ref{fig1} (b) and larger ones are 
shown in Fig. \ref{fig1} (c). In addition, we
plot the corresponding obtained OCT pulses (black dashed lines in lower
panels of Fig. (\ref{fig1})).

\begin{figure}[hbt]
\includegraphics[width=1.0\linewidth]{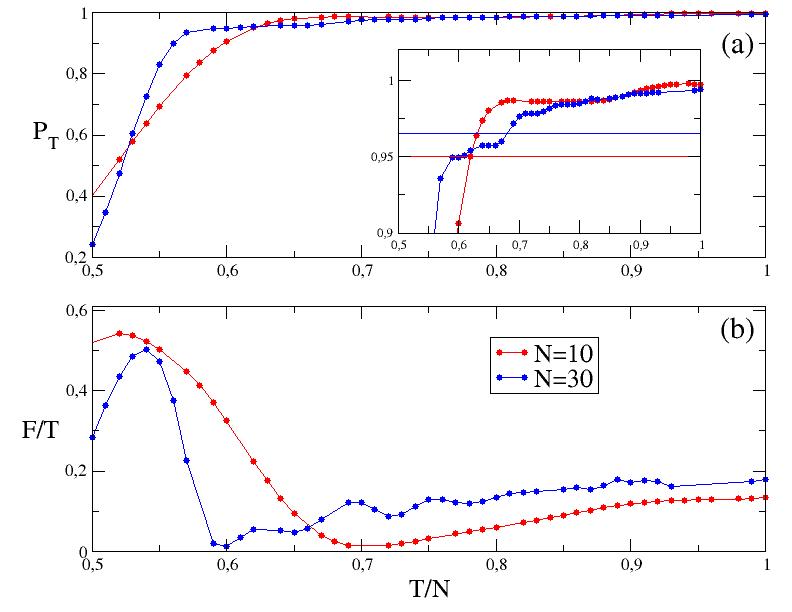}
\caption{ \label{fig2} In panel (a) we plot target population as a
function of the normalized operation time for different chain lengths 
($N=10$ in red and $N=30$ in blue) obtained with OCT pulses. In panel 
(b) we observe the fluence of the OCT pulses as a function of the normalized
 operation time.
}
\end{figure}

We can see more details related to the OCT dynamics in Fig. \ref{fig2}. Panel 
(a)  shows the target probability obtained with OCT for two different 
chain sizes, $N=10$ in red circles and $N=30$ in blue circles, as a function 
of the normalized operation time. For short times, i.e. less than $\simeq 
0.65N$, the one actuator protocol fails to reach the desired 
target, obtaining target probabilities lower than the obtained using
a simple free evolution. This can be observed clearly in the inset with 
a continuous  straight  line with the corresponding color for each $N$. 
For $T\simeq 0.65N$, the one actuator OCT protocol reaches target probabilities 
similar to the one obtained in the peak of the free evolution. 
Remarkably, the reduced Fluence 
of the OCT actuator; which is defined as the Fluence scaled by the operation 
time $T$; presents a minimun near $T_{peak}$, as shown in the panel (b) of 
Fig. \ref{fig2}. For larger Control Times $T$, $F/T$ increases with a quasi 
linear dependence with $T$, which indicates the need of larger Fluence in 
order to obtain high target probabilities for larger times.

\begin{figure}[hbt]
\includegraphics[width=1.0\linewidth]{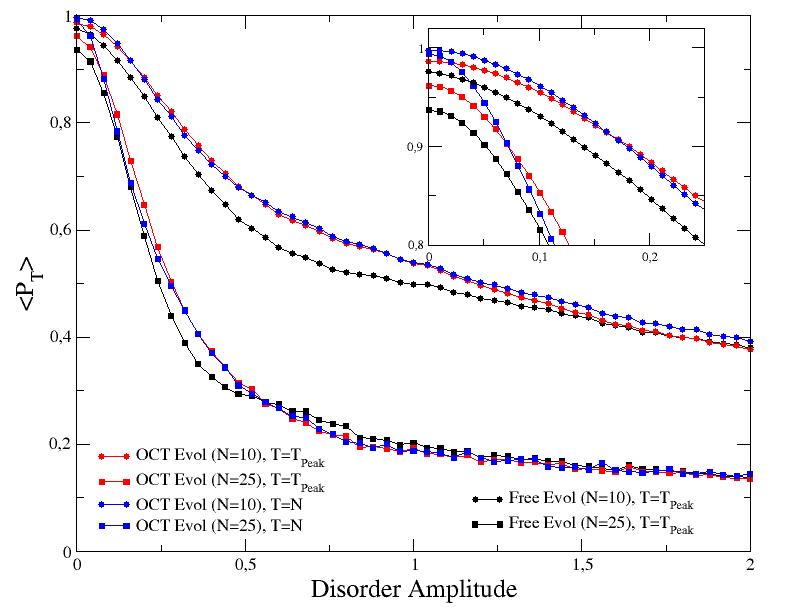}
\caption{ \label{figPtDis} Target population averaged over $2000$ 
realizations  as a function of Amplitude disorder for different chain 
lengths ($N=10$ with circles and $N=25$ with squares). In blue we observe 
target population for long pulses ($T_{OCT}=N$), in red for short pulses 
($T_{OCT}=T_{peak}$) and in black we see the free evolution peak. The
inset shows a zoom for small values of the amplitude disorder.
}
\end{figure}

Now, we will assess the effect of disorder over the controlled transfer 
protocol. Fig.~\ref{figPtDis} shows how the static disorder, as described in 
the 
Sec. \ref{sec:model-oct}, affects the target probability. The Figure also shows 
 the free evolution for comparison. Fig. \ref{figPtDis} 
show the averaged  target probability $\langle P_T\rangle$, which is obtained 
from over $M=2000$  realizations of the 
disorder. For each realization of the disorder the  target probability 
was obtained using the OCT protocol without disorder, and the averaged target 
probability is the unweighted average of all the realizations. The figure shows 
results
for two different length chains ($N=10$ and $N=25$), choosing two values of 
$T_{OCT}$, namely $T_{OCT}\simeq T_{peak}$ and $T_{OCT}=N$, and the averaged 
target probability is plotted versus the 
disorder amplitude $A$. 
As expected with increasing values of $A$ the target probability decreases.

The inset in the figure shows that the better yield for very small disorder 
amplitude is obtained by using the optimized pulse for the larger operation 
time. However, a very small disorder amplitude destroys the improvement 
obtained 
by increasing the operation time. For amplitudes $A>0.1$ there is no 
difference between optimized evolution with $T_{OCT}=T_{peak}$ or 
$T_{OCT}=N$. This is not the case when we compare with the free evolution. We 
need to go to disorder amplitudes $A>1.5$ in order to obtain similar results 
using a free evolution strategy or employing an optimized pulse.

\begin{figure}[hbt]
\includegraphics[width=1.1\linewidth]{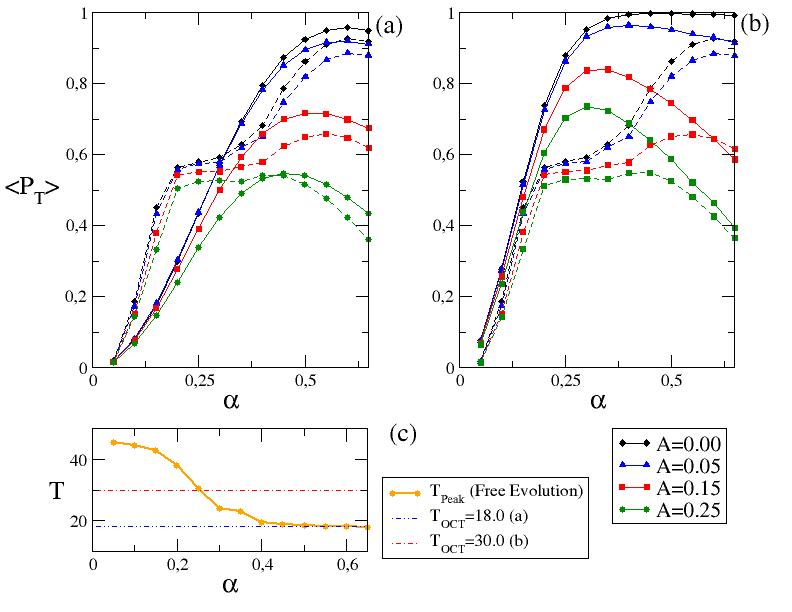}
\caption{Upper panels show the averaged target population over $2000$
realizations as a function of the $N=30$ chain parameter $\alpha$ for
OCT pulses (Continuous lines) and free evulotion (dasshed
lines) for different disorder amplitudes. Panel (a) shows results for small 
operation time ($T_{OCT}=T_{peak}$ corresponding to the optimal $\alpha$, 
for $N=30$ we have $T\simeq 18$ a.u.) and panel (b) shows results for long 
operation time ($T_{OCT}=N$). Panel (c) shows $T_{peak}$ as a
function of the $N=30$ chain parameter $\alpha$.
}\label{figPtDis2}
\end{figure}

A simple way  to improve the transmission using the free evolution
strategy is to optimize the exchange coupling of the extremes $\alpha$
as was shown in several works \cite{Zwick1,Zwick2,Zwick3}. In
Fig. \ref{figPtDis2} we plot the transfer probability for the free
evolution at $T=T_{peak}$ as a function of $\alpha$. We also plot
the target probability for the optimized dynamics for different
operation times. All curves show a maximum close to $\alpha=0.6$. This
result was known for the free evolution case. In this figure we observe an
important advantage in the optimized evolution and the impaact of
using large operation times. It is clear that, for $T_{OCT}=N$ we can
use a wide range of $\alpha$ values, more than if we employ
$T_{OCT}=T_{peak}$ and much more than in the free evolution case. It is 
worth to mention that, for the free evolution, exists a spurious shoulder 
related to peak times much larger than $N$ as we can observe in 
Fig. \ref{figPtDis2} (c).

\begin{figure}[hbt]
\includegraphics[width=1.0\linewidth]{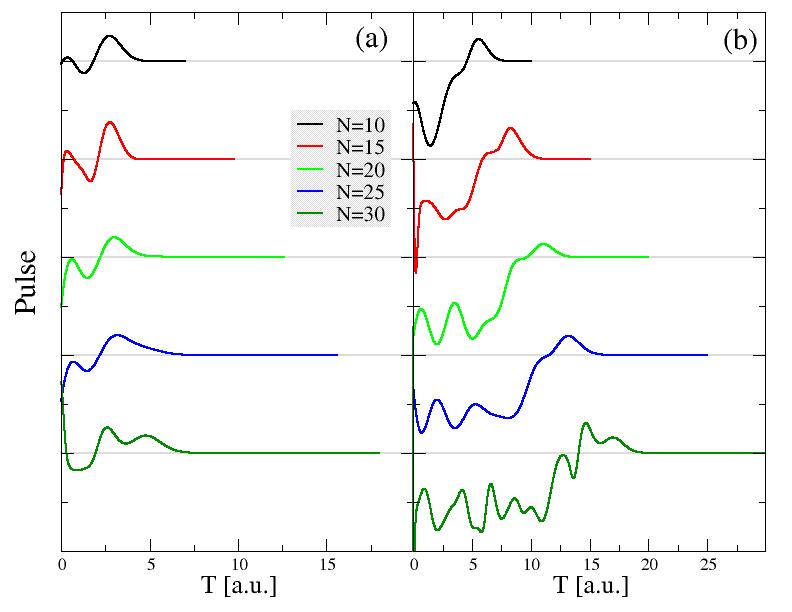}
\caption{OCT pulses as a function of Time for different chain lengths.
In (a) we plot short pulses ($T_{OCT}\simeq T_{peak}$) and in 
(b) we show the long pulses ($T_{OCT}=N$) }
\label{un-actuator-vs-n}
\end{figure}

Finally, Fig. \ref{un-actuator-vs-n} shows the amplitude of the control
pulse as a function of time, for different chain lengths.
As can be observed, the pulses are appreciably different from zero in almost 
the same time interval for different values of $N$. This statement seems
to be more accurate in the case of short pulses. In all the cases presented
here, these times are shorter than the characteristic times where the free 
evolution lead to a maximum in the population transfer and shorter than the 
operation time in the OCT dynamics.

Interestingly, Figure~\ref{un-actuator-vs-n} strongly suggest that the
control pulses of shorter chains can be used as initial pulses for the OCT
iterative procedure when dealing with larger chains. Moreover, it suggests that
a simpler analytical pulse can be constructed in terms of the natural
frequencies of the chain spectrum and some envelope function.

Optimal control theory, on the other hand, allows us to choose the
desired time at which the maximum transfer is achieved, so
synchronizing several signals to arrive at a given time is a certain possibility.
Panel (b) of Fig. \ref{un-actuator-vs-n} shows what happens if longer 
arrival ($O(N)$) times are chosen. The pulses are, again, fairly simple but 
the number of frequencies involved increases with the transfer time, this 
behavior is consistent with the control pulses found in the literature that 
used control times on the order of $O(N^2)$. In a sense, the high frequency 
pulses diminishes the net propagation speed of the excitation transfer process.

Here we want to return briefly to the results of Figure~\ref{fig1}  to offer a 
qualitative description of how the transmission process is carried out with 
control pulses like those that we have calculated.
In Fig.~\ref{fig1} (b) and (c) we can
appreciate that the target population remains null while the pulse is 
appreciably different from zero. When the pulse dies and the evolution
start to be a free evolution, the target population begins to rise. This can 
be understood as follows: the control pulse prepares a state that once it is 
prepared propagates along the spin chain until the population becomes 
localized at the other extreme of the chain. The pulse has an  appreciable 
amplitude while 
the preparation takes place, but after this time there is no need for further 
driving. Besides, since the information propagates at the 
Quantum Speed Limit, the actuator can not affect the propagation once the 
state is ``too far away''. Of course this semi-classical analysis should not be 
taken literally, but offers a good picture of the transfer process.

\section{Control with two actuators}\label{sec:dos-controles}

This Section is dedicated to analyze a different control strategy, using two 
actuators instead of just a single one, and assess if the physical traits found 
in the previous Section with respect to the control pulses, control time and 
robustness against static disorder are shared by more complex control 
strategies. To this end we implemented control equations where 
 the first and last exchange couplings are time-dependent control functions. 
Despite the excellent 
controllability and the high population in the last site of the chain achieved 
after the transfer process, 
we want to explore here if the use of one more actuator enhances the good 
properties or if the trade off between simplicity and gains is not favorable.

\begin{figure}[hbt]
\includegraphics[width=1.05\linewidth]{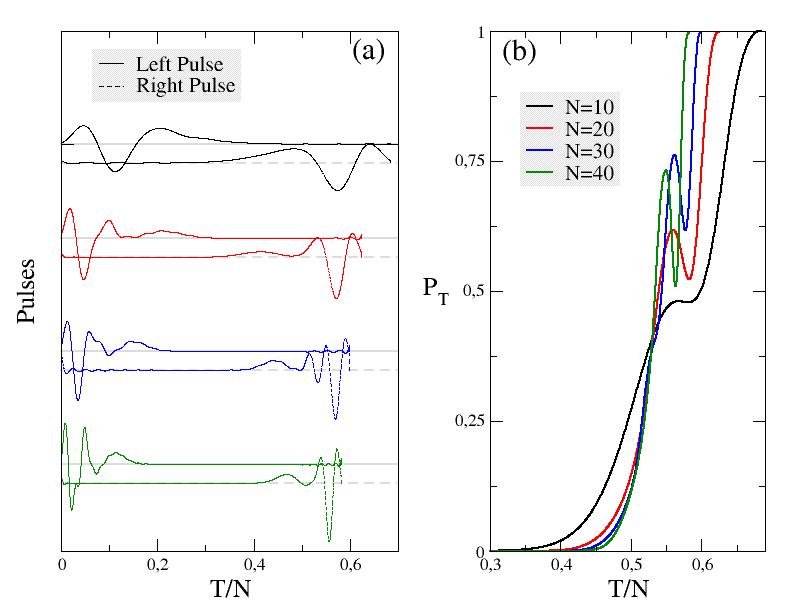}
\caption{OCT dynamics with Two Actuators. (a) Pulses for different
chain length as a function of the normalized time $T/N$. (b) Population
of the target state as 
a function of the normalized time obtained with the OCT pulses showed in panel
(b).}
\label{dos-actuator-resumen}
\end{figure}

Figure~\ref{dos-actuator-resumen} a) shows the control functions, as functions 
of the normalized time $T/N$, at both extremes of the chain and, from top 
to bottom, for chain lengths $N=10,20,30$ and $40$. The solid lines correspond 
to the control functions at the first exchange coupling and the dashed ones to 
the control functions at the last one. Both pulses, for a given chain length, 
were obtained as independent functions. Despite this, the reflection symmetry 
around one half of the normalized time is obvious. This result can be 
understood in terms of the reflection symmetry obeyed by the spin couplings 
that is characteristic of perfect state transfer schemes.  This could 
be a practical recipe to look for control pulses with two actuators, {\em i.e.} 
instead of looking for two independent control pulses it could be more simpler 
to look for ``reflected pulses'' to be applied at the opposed ends of the 
chain. At any rate, the results presented in what follows were obtained 
assuming that the two control pulses were independent. It is worth to point 
that, again, the pulses for different chain lengths are closely related and 
that the pulses of shorter chains can be used as the initial pulses of the OCT 
method to find the optimized pulses of longer chains.

Figure~\ref{dos-actuator-resumen} (b) shows the population at the last site of 
the chain obtained when the pulses shown in panel (a) control the dynamical 
behavior of the spin chain. The bigger difference with the rather monotonous 
behavior shown by the population growth when only one actuator is controlling 
the dynamics  is the ``shoulder'' near $0.5 T/N$ (see 
Figure~\ref{dos-actuator-resumen}).

\begin{figure}[hbt]
\includegraphics[width=1.0\linewidth]{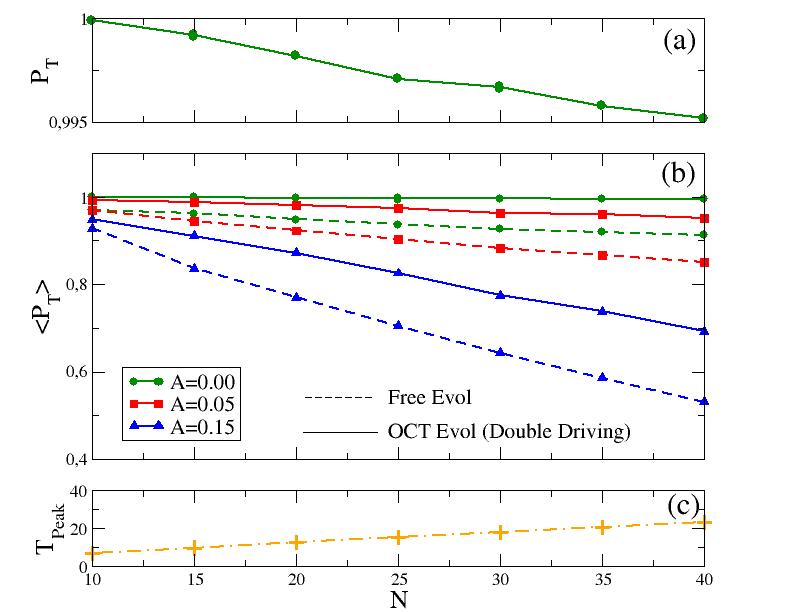}
	\caption{Disorder effect in the OCT evolution using two actuators. In
panel (a) we show the yield as a function of the length chain for short OCT 
pulses using two actuators. The operation time employed in the transfer process
$T_{OCT}=T_{peak}$ corresponding to he free evolution peak for the optimal 
$\alpha$ plotted in panel (c). In panel (b) we observe the effect of the 
disorder in the control operation
	}
\label{dos-actuator-resumen2}
\end{figure}

The robustness against static disorder for the two actuator case was analyzed 
in a similar way that  the one actuator case. Using the control pulses 
designed to maximize the population transfer in an ordered chain we studied the 
population transfer in chains with static disorder. So, the results shown 
in Fig. \ref{dos-actuator-resumen2} are averages over $M=2000$ realizations of 
the disorder. Panels (a), (b) and (c) of 
Fig. \ref{dos-actuator-resumen2} show the maximum 
population transfer achieved, the averaged population at peak time for several 
disorder strengths, and the peak time, respectively, as functions of the chain 
length $N$. The data in Panel (c), show that the population transfer is 
achieved at peak times which change linearly when the chain length is 
increased.  Panel (b) shows that, as expected, the controlled two-actuator 
scheme is far more robust against disorder than the free evolution of the 
border optimized chains, the population transfer corresponding to the 
two-actuators scheme (dots joines by solid lines) is always above the free 
evolution data (dots joined by dashed lines) and the decaying rate with the 
chain length is smaller. 
On the other hand, as shown in panel (a)  
the population 
transfer for the undisturbed chain is above $0.995$ even for $N=40$, which is 
excellent. It is important to remark that this level of control is achieved 
with low fluence pulses.

\begin{figure}[hbt]
\includegraphics[width=1.0\linewidth]{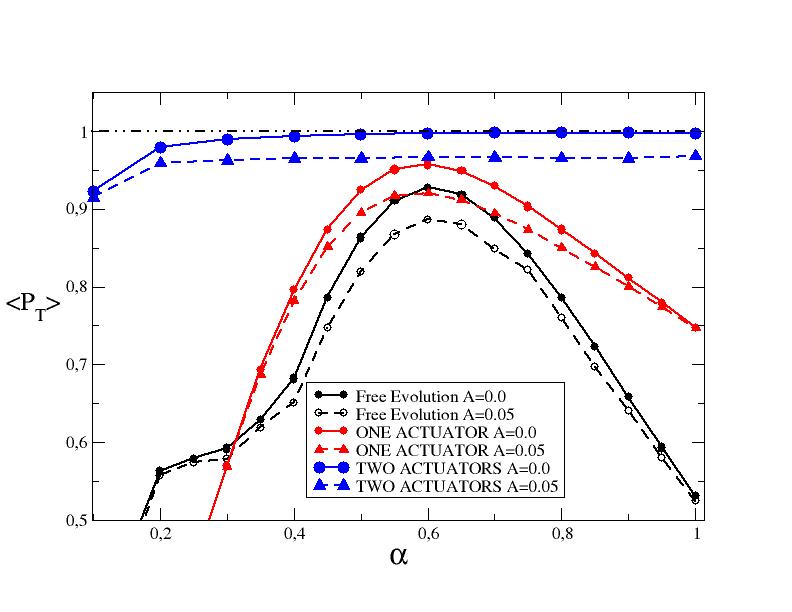}
\caption{ \label{dos-actuator-resumen3}
Averaged target population over $2000$ realizations as a function of the 
$N=30$ chain length $\alpha$ parameter for one actuator OCT pulses 
(red lines), for two actuators OCT pulses (blue lines) and for free evolution 
(black lines) with different disorder amplitudes. 
The result presented here are obtained for small operation 
time ($T_{OCT}=T_{peak}$ corresponding to the optimal $\alpha$, for $N=30$ 
we have $T_{OCT}\simeq 18$ a.u.) 
}
\end{figure}

The comparison between the one and two-actuators
schemes performance is shown in Fig. \ref{dos-actuator-resumen3}. This 
figure  shows the population transfer, with and 
without disorder, achieved with 
one and two-actuators schemes, for $N=30$. The free evolution is also depicted 
for comparison. The population transfer are shown, for all the cases, 
as functions of the border parameter $\alpha$. As shown by the Figure, the one 
actuator scheme and the free evolution depend strongly on this parameter, 
showing that it must be precisely tuned to attain a high population transfer. 
In contradistinction, the two actuator scheme is almost independent from it. As 
can be observed the two-actuator control achieves better population transfer 
than the one-actuator, even when the former is applied to disordered 
chains ($A=0.05$) and the later is applied to chains without disorder, except 
for a small interval of values of $\alpha$ near $\alpha=0$. This observation 
shows that the two-actuator scheme clearly outperforms the one-actuator one, of 
course  barring any difficulties associated to the fact that the population 
transferred must be checked in one of the spins over which one of the actuators 
is acting.

\section{Conclusions}

The results presented in the previous Sections show that, using OCT, it is 
possible to design smooth, simple and reliable control pulses, that transfer 
population from one extreme of a $XX$ spin chain to the other. The transfer can 
be achieved choosing the arrival time, as long as this time is longer or 
similar to the time compatible with the QSL, $T_{QSL}\approx N/2$. Moreover, in 
the one-actuator scheme,  if the arrival time is chosen similar to $T_{QSL}$ 
the pulses will have very good properties as low fluence and the switch-off can 
be made smoothly and without a precise time control since the last part of the 
pulse has a very small amplitude and hardly affects the population transfer.

In the case of the two-actuator scheme, even for moderate chain lengths, the 
pulses act  in (practically) different time intervals so, despite that an 
actuator has been added, there is no need of a precise timing between the two 
pulses. If this were not the case probably the need of a highly precise timing 
between the pulses could spoil  the higher population transfer gained using two 
actuators instead of just one. If a two-actuator strategy were not feasible, 
tuning the value of the last exchange coupling constant contributes effectively 
to increase the population transfer.

It is necessary to remark the incredible robustness against the $\alpha$ 
parameter change of the two actuator protocol. For small operation times
the difference with the one actuator protocol is huge. As we observe in Fig. 
\ref{dos-actuator-resumen3} the one actuator protocol
and the free evolution have a maximum for $\alpha\simeq 0.6$ while
the double driving protcol shows an almost constant behaviour with 
values of the yield higher in all the range of the $\alpha$ parameter
showed here. In the case of long operation times the differences between 
both protocols is not so important for values of $\alpha\geq 0.3$ as
shown in Fig. \ref{dos-actuator-resumen3} and Fig. \ref{figPtDis2}.

Most studies that can be found in the literature consider that the control is 
exerted by external magnetic fields, which is natural for Magnetic Nuclear 
Resonance scenarios in liquids. For solid state settings, as quantum dots or 
superconducting qubits chains,  the manipulation of the couplings between 
adjacent sites seems more achievable and faster. 

Our findings also support the conclusion that systems with a QSL for the 
propagation of information should be controllable at times compatibles with 
such limit, in this sense we are working with other quantum spin chains to 
verify this. In particular we are interested if this limit can be achieved 
controlling exchange couplings or magnetic fields within schemes with at most 
two actuators acting over fixed sites.

{\em Acknowledgments}
A. F., and S. S. G. acknowledge financial support from
CONICET (PIP11220150100327, PUE22920170100089CO). O.O. acknowledges partial 
financial support from CONICET and SECYT-UNC. A.F. thanks the warm hospitality 
of the FAMAF, National University of C\'ordoba where this collaboration 
initiated.

\end{document}